\documentclass[reprint,aps,prd,superscriptaddress,nofootinbib]{revtex4-2}

\usepackage{dcolumn}
\usepackage{bm}

\usepackage{graphicx}
\usepackage{subcaption}
\usepackage{array}

\usepackage{booktabs}
\usepackage{multirow}

\usepackage{siunitx}
\usepackage{xcolor}
\usepackage{ulem}
\usepackage[colorlinks=true, allcolors=blue]{hyperref}

\usepackage[utf8]{inputenc}
\usepackage[T1]{fontenc}
\usepackage{amsmath,amssymb}

\usepackage{tikz}
\usetikzlibrary{shapes.misc, positioning, arrows.meta}

\begin{document}

\preprint{APS/123-QED}
\title{Scalar Non-standard Interactions in Type-II Seesaw }

\author{Bhaswati Kalita}
\email{bhaswatik@tezu.ernet.in}

\author{Mrinal Kumar Das}
\email{mkdas@tezu.ernet.in}

\affiliation{Department of Physics, Tezpur University, Napaam, Sonitpur, Assam 784028, India
}

\begin{abstract}

We study the theoretical origin of scalar non-standard interactions (SNSI) in an \(A_4\)-symmetric Type-II seesaw model. The model contains an \(SU(2)_L\) scalar triplet \(\Delta\) and a real scalar singlet \(S\). After electroweak symmetry breaking, their CP-even neutral fluctuations mix to form two physical scalar states \(\phi_1\) and \(\phi_2\). These physical scalars couple to both neutrinos and charged leptons, establishing a connection between the neutrino sector and ordinary matter through scalar interactions. This induces a matter-dependent correction to the Type-II seesaw neutrino mass matrix. The SNSI contribution, governed by the relevant Yukawa couplings and the SNSI coefficient, can enhance or suppress the neutrino masses relative to their vacuum values. The model provides a simple and consistent description of how scalar mixing can give rise to SNSI and connects neutrino interactions in matter with the underlying structure of the scalar sector 
\end{abstract}

\maketitle

\noindent\textit{Introduction ---} The observation of tiny neutrino masses and flavor mixing reveals an important limitation of the Standard Model (SM) \cite{bilenky2015neutrino}, indicating the need for physics beyond its minimal framework. Explaining the origin of these properties remains a fundamental question in particle physics and continues to motivate the development of new theoretical frameworks for physics beyond the Standard Model (BSM) \cite{crivellin2024anomalies}.
Among the various mechanisms, seesaw mechanisms \cite{cai2018lepton,king2004neutrino,miranda2016neutrino,king2016neutrino,das2020constraining}  provide a natural and elegant explanation for the smallness of neutrino masses  
\cite{smirnov2009neutrino,gustafsson2020lepton,lei2026neutrino,kaladharan2024unified} , by extending the SM with new particles and interactions . Within this framework, the type-II seesaw mechanism \cite{melfo2012type,gu2006minimal,barrie2022type,nomura2025type,das2012quasidegenerate,zhou2022electroweak,arun2022testing}  is particularly appealing because it generates neutrino masses through an \(SU(2)_L\) scalar triplet\cite{chiang2012search,arhrib2011higgs,ashanujjaman2022revisiting} whose neutral component acquires a small vacuum expectation value. It also provides a suitable framework for studying additional interactions involving neutrinos. A further important question is the origin of the observed pattern of neutrino mixing \cite{smirnov2014theory,gonzalez2003neutrino,petcov2005towards}. Discrete flavor symmetries \cite{altarelli2010discrete,king2013neutrino,ishimori2010non,chauhan2024phenomenology} offer a natural framework for organizing the lepton flavors and constraining the structure of the neutrino mass matrix \cite{ma2006neutrino}. Among these, the \(A_4\) symmetry \cite{kumar2025neutrino} is particularly suitable for neutrino models, as its triplet representation naturally accommodates the three generations of leptons. At the same time, neutrinos may interact with matter through processes beyond the standard weak interaction \cite{malinsky2009nonstandard}. Scalar non-standard interactions (SNSI) \cite{ge2019scalar,dutta2024non,dutta2026scalar,medhi2025scalar,babu2020neutrino,verma2026probing,sarker2024impact} can arise when a scalar mediator couples to neutrinos and matter. Such interactions can affect neutrino propagation in matter and may lead to modifications in neutrino mass and mixing. 

In this Letter, we bring these ideas together within an \(A_4\)-symmetric Type-II seesaw framework by introducing an additional real scalar singlet \(S\). The presence of \(S\) provides a simple way to connect the scalar sector responsible for neutrino mass generation with interactions involving ordinary matter. After electroweak symmetry breaking (EWSB) \cite{chen2012predictions}, the CP-even neutral fluctuations of the Higgs triplet \(\Delta\) and the scalar singlet \(S\) mix with each other, leading to two physical scalar states, \(\phi_1\) and \(\phi_2\). These scalar states can couple to both neutrinos and charged leptons, thereby providing a link between the neutrino and matter sectors. Their exchange gives rise to SNSI \cite{pusty2025impact,denton2025solar}, providing a natural extension of the Type-II seesaw framework.

\noindent\textit{The Model ---}    In this work, we consider an \(A_4\)-symmetric Type-II seesaw framework extended by a real scalar singlet \(S\). The model combines the flavour structure of the \(A_4\) symmetry with an extended scalar sector containing the \(SU(2)_L\) triplet and the singlet scalar. SNSI arises from the mixing of the CP-even singlet fluctuation \(s\) and the CP-even neutral triplet fluctuation \(\delta_{1R}\). The resulting scalar mass eigenstates couple to both neutrinos and charged leptons, inducing an effective scalar-mediated neutrino–electron interaction in matter. The main purpose of this model is to study how SNSI can arise theoretically within the considered neutrino mass model and to explore their phenomenological consequences. The field content and their transformation properties under the gauge symmetry $SU(2)_L \times U(1)_Y$ and the discrete flavour symmetry $A_4$ are shown in Table I. The three left-handed lepton fields $L=(L_e, L_\mu, L_\tau)$ transform as doublets under \(SU(2)_L\) and triplet under $A_4$. The right-handed charged leptons \(e_R\), \(\mu_R\), and \(\tau_R\) transform as singlets under \(SU(2)_L\) and as \(1\), \(1^{\prime\prime}\), and \(1^\prime\) under \(A_4\), respectively. The Higgs fields \(H=(H_1,H_2,H_3)\) form a triplet under \(A_4\), with each \(H_i\) transforming as an \(SU(2)_L\) doublet. Similarly, \(\Delta=(\Delta_1,\Delta_2,\Delta_3)\) forms an \(A_4\) triplet, with each \(\Delta_i\) transforming as an \(SU(2)_L\) triplet with hypercharge \(Y=+1\). In addition, \(S\) is a real scalar singlet under \(SU(2)_L\) and \(A_4\).

\begin{table}[ht]
\centering
\renewcommand{\arraystretch}{1.4}
\begin{tabular}{|c|c|c|c|c|}
\hline
\textbf{Particle} & $\mathbf{SU(2)_L}$ & $\mathbf{U(1)_Y}$ & $\mathbf{A_4}$ \\
\hline

$L=(L_e,L_\mu,L_\tau)$ & $2$ & $\dfrac{-1}{2}$ & $\mathbf{3}$ \\
\hline

$e_R$ & $1$ & $-1$ & $\mathbf{1}$ \\
\hline

$\mu_R$ & $1$ & $-1$ & $\mathbf{1''}$ \\
\hline

$\tau_R$ & $1$ & $-1$ & $\mathbf{1'}$ \\
\hline

$H=(H_1,H_2,H_3)$ & $2$ & $\dfrac{1}{2}$ & $\mathbf{3}$ \\
\hline

$\Delta=(\Delta_1,\Delta_2,\Delta_3)$ & $3$ & $+1$ & $\mathbf{3}$ \\
\hline

$S$ & $1$ & $0$ & $\mathbf{1}$ \\
\hline

\end{tabular}
\caption{Particle content and transformation properties under the gauge symmetry $SU(2)_L \times U(1)_Y$ and the discrete flavour symmetry $A_4$.}
\label{tab:field_content}
\end{table}

The complete Lagrangian of this model can be written as - 
\[
\mathcal{L}
=
\mathcal{L}_{\mathrm{K.E.}}
+
\mathcal{L}_{\mathrm{P.E.}}
+
\mathcal{L}_{\mathrm{Yukawa}}
\]
Here, \(\mathcal{L}_{\mathrm{K.E.}}\), \(\mathcal{L}_{\mathrm{P.E.}}\), and \(\mathcal{L}_{\mathrm{Yukawa}}\) represent the kinetic, potential, and Yukawa sectors of the model, respectively.

The complete Kinetic Energy of the model can be written as - 
\begin{equation}
{\mathcal{L}_{\rm kin}=\mathcal{L}_{\rm gauge}+
\mathcal{L}_{\rm fermion}+\mathcal{L}_H+\mathcal{L}_\Delta
+\mathcal{L}_S}
\end{equation}
The explicit form of the kinetic energy, including the contributions from all the fields in the model is given by
\begin{equation}
\begin{aligned}
\mathcal{L}_{\rm kin} ={}&
-\frac{1}{4}W^a_{\mu\nu}W^{a\mu\nu}
-\frac{1}{4}B_{\mu\nu}B^{a\mu\nu}
+i\sum_{\alpha=1}^{3}
\overline{L_\alpha}\gamma^\mu D_\mu L_\alpha  \\
&+ i\overline{e_R}\gamma^\mu D_\mu e_R
+ i\overline{\mu_R}\gamma^\mu D_\mu\mu_R
+ i\overline{\tau_R}\gamma^\mu D_\mu\tau_R \\
&+ \sum_{i=1}^{3}(D_\mu H_i)^\dagger(D^\mu H_i)
+ \sum_{i=1}^{3}
\operatorname{Tr}\!\left[
(D_\mu\Delta_i)^\dagger(D^\mu\Delta_i)
\right] \\
&+ \frac{1}{2}\partial_\mu S\,\partial^\mu S 
\end{aligned}
\end{equation}

where,
\begin{equation}
D_\mu L_{\alpha}
=
\left(
\partial_\mu
- i g \frac{\sigma^a}{2} W_\mu^a
+ i \frac{g'}{2} B_\mu
\right) L_{\alpha} 
\end{equation}
\begin{equation}
D_\mu \ell_R
=
\left(
\partial_\mu+i g' B_\mu
\right)\ell_R
\qquad
\ell_R=e_R,\mu_R,\tau_R 
\end{equation}
\begin{equation}
D_\mu H_i
=
\left(
\partial_\mu
- i g \frac{\sigma^a}{2} W_\mu^a
- i \frac{g'}{2} B_\mu
\right) H_i 
\end{equation}
\begin{equation}
D_\mu \Delta_i
=
\partial_\mu \Delta_i
- i g
\left[
\frac{\sigma^a}{2} W_\mu^a,\Delta_i
\right]
- i g' B_\mu \Delta_i
\end{equation}

The complete potential energy of the model can be written as-

\begin{equation}
\begin{aligned}
V(H,\Delta,S)={}&
-m_H^2(H^\dagger H)_{\mathbf{1}}
-m_\Delta^2\operatorname{Tr}
(\Delta^\dagger\Delta)_{\mathbf{1}}
\\
&+\frac{1}{2}\mu_S^2S^2
+\frac{\mu_3}{3}S^3
+\frac{\lambda_S}{4}S^4
\\
&+\lambda_{H1}\big[(H^\dagger H)_{\mathbf{1}}\big]^2
+\lambda_{H2}(H^\dagger H)_{\mathbf{1}'}
(H^\dagger H)_{\mathbf{1}''}
\\
&+\lambda_{H3}
\big[(H^\dagger H)_{\mathbf{3}_s}\cdot
(H^\dagger H)_{\mathbf{3}_s}\big]
\\
&+\lambda_{\Delta1}
\big[\operatorname{Tr}(\Delta^\dagger\Delta)_{\mathbf{1}}\big]^2
\\
&+\lambda_{\Delta2}
\operatorname{Tr}(\Delta^\dagger\Delta)_{\mathbf{1}'}
\operatorname{Tr}(\Delta^\dagger\Delta)_{\mathbf{1}''}
\\
&+\lambda_{\Delta3}
\big[\operatorname{Tr}(\Delta^\dagger\Delta)_{\mathbf{3}_s}
\cdot
\operatorname{Tr}(\Delta^\dagger\Delta)_{\mathbf{3}_s}\big]
\\
&+\lambda_1(H^\dagger H)_{\mathbf{1}}
\operatorname{Tr}(\Delta^\dagger\Delta)_{\mathbf{1}}
\\
&+\lambda_2(H^\dagger H)_{\mathbf{1}'}
\operatorname{Tr}(\Delta^\dagger\Delta)_{\mathbf{1}''}
\\
&+\lambda_3(H^\dagger H)_{\mathbf{1}''}
\operatorname{Tr}(\Delta^\dagger\Delta)_{\mathbf{1}'}
\\
&+\lambda_4(H^\dagger H)_{\mathbf{3}_s}\cdot
\operatorname{Tr}(\Delta^\dagger\Delta)_{\mathbf{3}_s}
\\
&+\lambda_6(H^\dagger\Delta\Delta^\dagger H)_{\mathbf{1}}
+\mu_{HS}S(H^\dagger H)_{\mathbf{1}}
\\
&+\lambda_{HS}(H^\dagger H)_{\mathbf{1}}S^2
\\
&+\mu_{\Delta S}S
\operatorname{Tr}(\Delta^\dagger\Delta)_{\mathbf{1}}
+\lambda_{\Delta S}
\operatorname{Tr}(\Delta^\dagger\Delta)_{\mathbf{1}}S^2
\\
&+\mu_{\Delta}
(H^T i\sigma_2\Delta^\dagger H)_{\mathbf{1}}
\\
&+\lambda_{\rm mix}S
(H^T i\sigma_2\Delta^\dagger H)_{\mathbf{1}}
+\mathrm{h.c.}
\end{aligned}
\label{eq:scalar_potential}
\end{equation}

The Yukawa sector of the model consists of different terms which are responsible for generating the charged-lepton masses, the neutrino mass through the Type-II seesaw mechanism, and the scalar-mediated interactions leading to the SNSI. Thus, the complete Yukawa Lagrangian can be written as
$$
\mathcal{L}_{Y}= \mathcal{L}_{Y}^{\ell}+ \mathcal{L}_{Y}^{\nu}
+ \mathcal{L}_{Y}^{S}$$

where \(\mathcal{L}_{Y}^{\ell}\), \(\mathcal{L}_{Y}^{\nu}\), and \(\mathcal{L}_{Y}^{S}\) represent the charged-lepton, neutrino, and scalar-mediated Yukawa interactions, respectively.

The charged-lepton masses arise from the \(A_4\)-invariant Yukawa interactions. The corresponding Yukawa Lagrangian is given by
\begin{equation}
\label{eq:8}
\mathcal{L}_{\ell}
= y_e(\overline{L}H)_{1}e_R + y_\mu(\overline{L}H)_{1'}\mu_R
+ y_\tau(\overline{L}H)_{1''}\tau_R +h.c.
\end{equation}

Using Eq. \eqref{eq:8}, the charged lepton mass matrix can be obtained. It is expressed as given in Eq. \eqref{eq:9}.
\begin{equation}
\label{eq:9}
M_\ell=\frac{v}{\sqrt{2}}
\begin{pmatrix}
y_e&0&0\\
0&y_\mu&0\\
0&0&y_\tau
\end{pmatrix}
\end{equation}
In addition to the standard Yukawa interaction, the singlet scalar
$S$ couples to the charged-lepton sector through the effective
dimension-five operators
\begin{equation}
\mathcal{L}_{S\ell}
=
\frac{g_e}{\Lambda}S\,\overline{L}H e_R
+
\frac{g_\mu}{\Lambda}S\,\overline{L}H\mu_R
+
\frac{g_\tau}{\Lambda}S\,\overline{L}H\tau_R
+h.c.
\end{equation}

Considering a non-zero vacuum expectation value of the singlet scalar

\begin{equation}
S(x)=v_S+s(x)
\end{equation}

the above interaction becomes

\begin{align*}
\mathcal{L}_{S\ell}= \frac{g_e}{\Lambda}(v_S+s)\overline{L}He_R
+\frac{g_\mu}{\Lambda}(v_S+s)\overline{L}H\mu_R
\\
+
\frac{g_\tau}{\Lambda}(v_S+s)
\overline{L}H\tau_R
+h.c.
\end{align*}

After EWSB, this gives two contributions:

\begin{equation}
\mathcal{L}_{S\ell}
=
\mathcal{L}_{mass}^{S}
+
\mathcal{L}_{int}^{S}
\end{equation}

where the singlet VEV contribution modifies the charged-lepton masses,
\begin{equation}
\label{eq:13}
\mathcal{L}_{mass}^{S}
=
m_e^{S}\overline{e_L}e_R
+
m_\mu^{S}\overline{\mu_L}\mu_R
+
m_\tau^{S}\overline{\tau_L}\tau_R
+h.c.
\end{equation}
with
\begin{equation}
m_\alpha^{S}
=
\frac{g_\alpha v v_S}{\sqrt{2}\Lambda}
\qquad
\alpha=e,\mu,\tau .
\end{equation}

The physical singlet fluctuation $s(x)$ gives the interaction with charged leptons,
\begin{equation}
\label{eq:15}
\mathcal{L}_{s\ell}
=
g'_e s\overline{e_L}e_R
+
g'_\mu s\overline{\mu_L}\mu_R
+
g'_\tau s\overline{\tau_L}\tau_R
+h.c.
\end{equation}

where

\begin{equation}
g'_\alpha
=
\frac{g_\alpha v}{\sqrt{2}\Lambda}
\qquad
\alpha=e,\mu,\tau .
\end{equation}

From the Lagrangian Eq. \eqref{eq:13}, the charged-lepton mass matrix can be written as given in Eq. \eqref{eq:17}.

\begin{equation}
\label{eq:17}
M_\ell^{(S)}
=
\frac{vv_S}{\sqrt{2}\Lambda}
\begin{pmatrix}
g_e&0&0\\
0&g_\mu&0\\
0&0&g_\tau
\end{pmatrix}
\end{equation}

The singlet VEV $v_S$ therefore modifies the charged-lepton masses, while the physical scalar field $s(x)$ remains responsible for the scalar-mediated interactions relevant for SNSI. 
Therefore, the complete charged-lepton mass matrix receives
contributions from both the usual Yukawa interaction and the singlet VEV contribution:

Hence, the total charged-lepton mass matrix takes a diagonal form, given in  Eq. \eqref{eq:18},

\begin{equation}
\label{eq:18}
M_\ell=
\frac{v}{\sqrt{2}}
\begin{pmatrix}
y_e+\frac{g_ev_S}{\Lambda}&0&0\\[2mm]
0&y_\mu+\frac{g_\mu v_S}{\Lambda}&0\\[2mm]
0&0&y_\tau+\frac{g_\tau v_S}{\Lambda}
\end{pmatrix}
\end{equation}

The light neutrino masses arise from the coupling of the lepton doublets to the scalar triplet. This interaction generates neutrino masses, and the relevant interaction term is
\begin{equation}
\mathcal{L}_{\nu} = Y_{\Delta}\left(L^{T} C\, i\sigma_{2}\, \Delta\, L\right)_{\mathbf{1}}+\mathrm{h.c.}
\end{equation}

Using the \(A_4\) product rules in the Altarelli–Feruglio basis, the Lagrangian can be expanded as 
\begin{align*}
\mathcal{L}_{\nu} =  & \left( \nu_e^T C \nu_e - \dfrac{\nu_\mu^T C \nu_\tau}{2} - \dfrac{\nu_\tau^T C \nu_\mu}{2} \right)\Delta_{1}^{0} \\
+ & \left( {\nu_\tau^T C \nu_\tau} - \dfrac{\nu_e^T C \nu_\mu}{2} - \dfrac{\nu_\mu^T C \nu_e}{2}\right)\Delta_{2}^{0} \\
+ & \left( \nu_\mu^T C \nu_\mu - \dfrac{\nu_e^T C \nu_\tau }{2}- \dfrac{\nu_\tau^T C \nu_e }{2}\right)\Delta_{3}^{0} + h.c.
\end{align*}

\begin{equation}
\Delta_i^{0} = \frac{1}{\sqrt{2}} (v_{\Delta} + \delta_{iR})
\end{equation}

where $v_{\Delta}$ denotes the vacuum expectation value generating the neutrino mass matrix, and $\delta_{iR}$ represents the CP-even scalar fluctuation of the triplet field.
The neutrino mass matrix $M_{\nu}$, arising from the vacuum expectation value $v_{\Delta}$ of the Higgs triplet $\Delta$, is given by:
\begin{equation}
 M_{\nu}^{\text{II}} = \frac{v_{\Delta}}{2\sqrt{2}} 
\begin{pmatrix} 
2Y_1 & -Y_2 & -Y_3 \\ 
-Y_2 & 2Y_4 & -Y_5 \\ 
-Y_3 & -Y_5 & 2Y_6 
\end{pmatrix}
\end{equation}
The scalar potential of the model is given above. To realize SNSI, we focus on the interaction term $$\lambda_{mix}S\left(H^{T}i\sigma_{2}\Delta^{\dagger}H\right)_{1}+\mathrm{h.c.}$$  which provides an explicit coupling between the singlet \(S\) and the triplet \(\Delta\). After EWSB, this interaction induces mixing between the CP-even neutral fluctuations of \(S\) and \(\Delta\), thereby providing the scalar mixing required for the SNSI mechanism.
After EWSB, the Higgs VEV alignment is considered as \(\langle H\rangle=(\dfrac{v}{\sqrt{2}},0,0)^T\). With the corresponding VEVs of the scalar triplet and singlet as \(\langle\Delta\rangle=(v_{\Delta1},v_{\Delta2},v_{\Delta3})\) and \(S(x)=v_{S}+s(x)\) , the mixing term $\lambda_{mix}S(H^T i\sigma_2\Delta^\dagger H)_1$ becomes
\begin{equation}
V_{\rm mix}= \frac{\lambda_{mix}v^2}{2\sqrt{2}}\left(v_Sv_\Delta+v_S\delta_{1R}+sv_\Delta
+s\delta_{1R}\right)
\end{equation}
After spontaneous symmetry breaking, the term proportional to $\lambda_{mix}$ generates several contributions involving the VEVs and scalar fluctuations. The terms proportional only to the VEVs or linear in \(s\) or \(\delta_{1R}\) contribute to the vacuum energy and the corresponding tadpole conditions. The relevant term for the physical scalar sector is the bilinear term
$$\frac{\lambda_{mix}v^2}{2\sqrt{2}}\,s\,\delta_{1R}$$
Here \(s\) and \(\delta_{1R}\) denote the CP-even neutral fluctuations of the singlet scalar \(S\) and the Higgs triplet \(\Delta\) respectively. This term mixes the two states through an off-diagonal entry in the scalar mass matrix and provides the key source of SNSI in the model.

We establish the relation between the gauge-interaction basis \((s,\delta_{1R})\) and the physical mass-eigenstate basis \((\phi_1,\phi_2)\) through a rotation by the mixing angle \(\theta\). The transformation is given by:
\begin{equation}
\begin{pmatrix} \phi_1 \\ \phi_2 \end{pmatrix} = \begin{pmatrix} \cos\theta & -\sin\theta \\ \sin\theta & \cos\theta \end{pmatrix} \begin{pmatrix} s \\ \delta_{1R} \end{pmatrix}
\end{equation}

\noindent Explicitly, the inverse transformation and the field relations are given by:

\begin{align}
s &= \cos\theta \phi_1 - \sin\theta \phi_2 \\
\delta_{1R} &= \sin\theta \phi_1 + \cos\theta \phi_2
\end{align}

\noindent Here, $\phi_1$ and $\phi_2$ represent the physical mass eigenstates, and $\theta$ is the mixing angle derived from the scalar potential parameters that diagonalize the mass-squared matrix of the $(s, \delta_{1R})$ system.
Since both the Higgs doublet $H$ and the scalar triplet $\Delta$ transform as triplets under $A_4$, the singlet contraction in the interaction \(\lambda_{\rm mix}S(H^Ti\sigma_2\Delta^\dagger H)_1\), together with the chosen Higgs VEV alignment, selects only the \(\Delta_1^0\) component. Consequently, only its \(\delta_{1R}\) component mixes with \(s\), providing the scalar-mixing mechanism responsible for the SNSI. The relevant neutrino interaction terms are

\begin{equation}
\label{eq:26}
\mathcal{L}_{\nu}(\delta_{1R}) = \frac{Y_1}{\sqrt{2}} \delta_{1R} \bar\nu_e^c \nu_e - \frac{Y_5}{2\sqrt{2}} \delta_{1R} \bar\nu_\mu^c \nu_\tau - \frac{Y_5}{2\sqrt{2}} \delta_{1R} \bar\nu_\tau^c \nu_\mu
\end{equation}

Substituting the expression for $\delta_{1R}$ into the neutrino interaction Lagrangian Eq. \eqref{eq:26} and grouping the terms by the physical scalars $\phi_1$ and $\phi_2$, we have $\mathcal{L}_{\nu} = \mathcal{L}_{\nu}(\phi_1) + \mathcal{L}_{\nu}(\phi_2)$, where:
\begin{align}
\mathcal{L}_{\nu}(\phi_1) &= \sin\theta \phi_1 \left[ \frac{1}{\sqrt{2}} Y_1 \bar{\nu}_e^c \nu_e - \frac{1}{2\sqrt{2}} Y_5 (\bar{\nu}_\mu^c \nu_\tau + \bar{\nu}_\tau^c \nu_\mu) \right] \\
\mathcal{L}_{\nu}(\phi_2) &= \cos\theta \phi_2 \left[ \frac{1}{\sqrt{2}} Y_1 \bar{\nu}_e^c \nu_e - \frac{1}{2\sqrt{2}} Y_5 (\bar{\nu}_\mu^c \nu_\tau + \bar{\nu}_\tau^c \nu_\mu) \right]
\end{align}

Similarly, substituting the mixing for $s$ into the charged-lepton interaction Lagrangian Eq. \eqref{eq:15} and 
separating the interactions mediated by the physical fields, $\mathcal{L}_{\ell} = \mathcal{L}_{\ell}(\phi_1) + \mathcal{L}_{\ell}(\phi_2)$, we find:
\begin{align}
\mathcal{L}_{\ell}(\phi_1) &= \cos\theta \phi_1 \left( g_e' \bar{e}_L e_R + g_\mu' \bar{\mu}_L \mu_R + g_\tau' \bar{\tau}_L \tau_R \right) \\
\mathcal{L}_{\ell}(\phi_2) &= -\sin\theta \phi_2 \left( g_e' \bar{e}_L e_R + g_\mu' \bar{\mu}_L \mu_R + g_\tau' \bar{\tau}_L \tau_R \right)
\end{align}

For ordinary terrestrial matter, electrons are abundant, while muons and taus are essentially absent.  Consequently, the expectation values of the fermion bilinears simplify to:
\begin{equation}
\langle \bar{\mu}\mu \rangle \simeq 0, \quad \langle \bar{\tau}\tau \rangle \simeq 0, \quad \text{while} \quad \langle \bar{e}e \rangle = N_e,
\end{equation}
The effective Lagrangian for $\phi_1$ is given by-
\begin{equation}
\mathcal{L}_{\text{SNSI}}(\phi_1) = \frac{2g'_e \sin2\theta }{m_{\phi_1}^2} N_e  \left[ \dfrac{Y_1 (\bar{\nu}_e^c \nu_e)}{\sqrt{2}} - \dfrac{Y_5 (\bar{\nu}_\mu^c \nu_\tau)}{2\sqrt{2}} - 
\dfrac{Y_5\bar{\nu}_\tau^c \nu_\mu)}{2\sqrt{2}} \right]
\end{equation}
 Similarly, the effective Lagrangian for $\phi_2$ can be given as- 

\begin{equation}
\mathcal{L}_{\text{SNSI}}(\phi_2) = \frac{-2g'_e \sin2\theta }{m_{\phi_2}^2} N_e  \left[ \dfrac{Y_1 (\bar{\nu}_e^c \nu_e)}{\sqrt{2}} - \dfrac{Y_5 (\bar{\nu}_\mu^c \nu_\tau)}{2\sqrt{2}} - \dfrac{Y_5\bar{\nu}_\tau^c \nu_\mu)}{2\sqrt{2}} \right]
\end{equation}
where $N_e$ represents the electron number density. 

 The total effective Lagrangian for the SNSI is given by-
\begin{equation}
\begin{aligned}
\mathcal{L}_{\text{SNSI}} = & 2g_e' N_e \sin2\theta  \left( \frac{1}{m_{\phi_1}^2} - \frac{1}{m_{\phi_2}^2} \right) \\
& \times \left[ \dfrac{Y_1}{\sqrt{2}} \bar{\nu}_e^c \nu_e - \frac{Y_5}{2\sqrt{2}} (\bar{\nu}_\mu^c \nu_\tau + \bar{\nu}_\tau^c \nu_\mu) \right]
\end{aligned}
\end{equation}

The SNSI induced effective mass matrix $\delta M$ is given by:
\begin{equation}
\begin{aligned}
\delta M ={}&
2g_e'N_e\sin2\theta
\frac{1}{2\sqrt{2}}
\left(
\frac{1}{m_{\phi_1}^2}
-\frac{1}{m_{\phi_2}^2}
\right)
\\[2mm]
&\times
\begin{pmatrix}
2Y_1 & 0 & 0\\
0 & 0 & -Y_5\\
0 & -Y_5 & 0
\end{pmatrix}
\end{aligned}
\label{eq:deltaM}
\end{equation}

The total neutrino mass matrix in matter $M_{\nu}$ is defined as the sum of the vacuum contribution from the Type-II Seesaw mechanism and the matter-induced correction due to SNSI.
\begin{equation}
M_{\nu} = M_{\nu}^{\text{II}} + \delta M
\end{equation}

Substituting the explicit matrix forms, we obtain:

\begin{equation}
M_{\nu} = \frac{v_{\Delta}}{2\sqrt{2}} 
\begin{pmatrix} 
2Y_1 & -Y_2 & -Y_3 \\ 
-Y_2 & 2Y_4 & -Y_5 \\ 
-Y_3 & -Y_5 & 2Y_6 
\end{pmatrix}
+ K
\begin{pmatrix} 
2Y_1 & 0 & 0 \\ 
0 & 0 & -Y_5 \\ 
0 & -Y_5 & 0 
\end{pmatrix}
\end{equation}
\begin{equation}
K =  2g' N_e \sin\theta \cos\theta\dfrac{1}{2\sqrt{2}} \left( \frac{1}{m_{\phi_1}^2} - \frac{1}{m_{\phi_2}^2} \right) 
\end{equation}
\noindent\textit{Summary and Discussions ---} We have developed an \(A_4\)-symmetric Type-II seesaw model to explore the origin of SNSI. In this model, SNSI modifies the conventional Type-II seesaw neutrino mass matrix through a matter-dependent contribution. The resulting mass matrix $M_{\nu}$ combines the vacuum neutrino mass term with the scalar interaction contribution. Depending on the strength of the SNSI contribution, which is determined by the relevant Yukawa couplings and SNSI coefficient, the different elements of $M_{\nu}$ can be enhanced or suppressed compared with their vacuum expectation values. This modification can have important consequences for neutrino phenomenology, including neutrino oscillations and other observables sensitive to the neutrino mass and mixing structure. Thus, the model provides a simple theoretical framework to investigate how SNSI can modify the neutrino mass sector and to study their possible signatures in different neutrino environments and future neutrino experiments.
\section*{acknowledgments}
BK acknowledges the financial support provided by the DST-INSPIRE Fellowship (DST/INSPIRE Fellowship/2024/IF240151). The authors sincerely thank Mr. Gourab Pathak and Mr. Pulakesh Borah for their valuable suggestions and helpful discussions throughout this work.
\label{LastBibItem}
\bibliographystyle{apsrev4-2}
\bibliography{apssamp}
\end{document}